\documentstyle[preprint,aps,epsf]{revtex}
\begin{document}
\draft
\preprint{}
\title{Hot Nuclear Matter in the Quark Meson Coupling Model}
\author{P. K. Panda$^{*\dagger}$, A. Mishra\footnote[1]{Alexander von Humboldt
Fellows (email: mishra@th.physik.uni-frankfurt.de)}$^\dagger$, 
J. M. Eisenberg$^\dagger$\footnote[4]{email: judah@giulio.tau.ac.il}
$^\ddagger$, W. Greiner$^\dagger$}
\address{$^\dagger$Institut f\"ur Theoretische Physik, 
J.W. Goethe Universit\"at, Robert Mayer-Stra{\ss}e 10,\\
Postfach 11 19 32, D-60054 Frankfurt/Main, Germany}
\address{$^\ddagger$ School of Physics and Astronomy,\\
Raymond and Beverly Sackler Faculty of Exact Sciences,\\
Tel Aviv University, 69978, Tel Aviv, Israel}
\maketitle
\begin{abstract}
We study here hot nuclear matter in the quark meson coupling (QMC) model
which incorporates explicitly quark degrees of freedom, with quarks coupled to
scalar and vector mesons. The equation of state of nuclear matter including 
the composite nature of the nucleons is calculated at finite temperatures.
The calculations are done taking into account the medium--dependent bag
constant. Nucleon properties at finite temperatures as calculated here 
are found to be appreciably different from the value at $T=0.$ 
\end{abstract}

\bigskip
\narrowtext

\newpage
\section{Introduction}

Quantum Hadrodynamics (QHD) is a model for the nuclear many-body problem
\cite{qhd,reinhard,serot2} describing nucleons interacting with scalar 
($\sigma$) and vector ($\omega$) mesons. In the mean-field approximation
for the meson fields, it has been shown that the ground-state properties 
of nuclear matter may be fitted by adjusting the scalar and vector couplings
\cite{walecka}. This meson field theory has quite successfully described the
properties of nuclear matter as well as finite nuclei using
hadronic degrees of freedom. The vacuum polarization corrections
arising from the nucleon fields \cite{chin} as well as the meson 
fields \cite{fox,sig} have also been considered to study nuclear matter.
This leads to a softer equation of state, giving a lower value of the
incompressibility than would be reached without quantum effects \cite{sig}.

While descriptions of the nuclear phenomena have been efficiently
formulated using hadronic degrees of freedom as in QHD, there have been 
interesting observations such as the EMC effect \cite{emc} which reveal 
the medium modification of the internal structure of the 
nucleon. This necessitates models that incorporate
quark--gluon degrees of freedom respecting the established
models based on the hadronic degrees of freedom. The Quark Meson
Coupling (QMC) model \cite{qmc} is a simple extension of QHD
incorporating the quark degrees of freedom. This model describes
nuclear matter with nucleons as non-overlapping MIT bags and the
quarks inside them couple to scalar and vector mesons. 
Though it is simple and attractive, this model predicts smaller
values for the scalar and vector potentials leading to a weaker
nucleon spin--orbit force in finite nuclei. The understanding of
the large and canceling values for the potentials, which has been
central to the success of nuclear phenomenology, has been attempted
recently through an analysis of a finite-density QCD sum-rule 
calculations \cite{sumrule} and is related to medium effects.
This is suggestive of including a medium-dependent bag constant 
in the QMC model \cite{jj,jjemc}. We intend to study here nuclear
matter using this variation of the QMC model at finite temperatures 
taking the medium-dependence of the bag parameters into account.

We organize the paper as follows: In section 2, 
we briefly recapitulate the QMC model for
nuclear matter at zero temperature \cite{jj}, and in section 3
we generalize it to finite temperature with a specific ansatz
for the density-dependent bag constant. Finally, in section 4,
we summarize the results as obtained in this model,
discuss its limitations, and present an outlook.

\section{QMC Model for Nuclear Matter}

We here briefly recapitulate the QMC model for nuclear matter
at zero temperature \cite{jj}. In this model, the nucleon in 
nuclear matter is assumed to be described by a static MIT bag in which quarks 
interact with the scalar ($\sigma$) and the vector ($\omega$) fields,
which are treated as classical in a mean field approximation.

The quark field $\psi_q(\vec r,t)$ inside the bag then satisfies the 
equation
\begin{equation}
\Big[i\gamma^\mu\partial_\mu-(m_q^0-g_\sigma^q\sigma)-g_\omega^q\omega\gamma^0
\Big]\psi_q(\vec r,t)=0,
\end{equation}
where $m_q^0$ is the current quark mass and $g_\sigma^q$ and $g_\omega^q$ 
are the quark couplings with the $\sigma$ and $\omega$ mesons.

The normalized ground state for a quark (in an s-state) in the bag
is given as
\begin{equation}
\psi_q(\vec r,t)=N \exp(-i\frac{\epsilon_q t}{R})
\left(
\begin{array}{c} j_0(xr/R)
\\ i\beta_q \vec \sigma \cdot \hat r j_1(xr/R)
\end{array}
\right) \frac{\chi_q}{\sqrt {4\pi}}
\end{equation}

Then the single particle quark energy in units of $R^{-1}$ is
\begin{equation}
\epsilon_q=\Omega_q+g_\omega^q\omega R, \quad
\beta_q=\sqrt{\frac{\Omega_q-R m_q^*}{\Omega_q+R m_q^*}},
\end{equation}
with
$\Omega_q=(x^2+R^2{m^*_q}^2)^{1/2}$;
$m^*_q=m^0_q-g_\sigma^q\sigma$ is the effective quark mass,
$R$ is the bag radius, $\chi_q$ is the quark spinor, and $N$ is the 
normalization factor.

The boundary condition at the bag surface is given by
\begin{equation}
i\gamma\cdot n \psi_q=\psi_q.
\end{equation}
This for the ground state reduces to 
\begin{equation}
j_0(x)=\beta_q j_1(x)
\label{x}
\end{equation}
which determines the dimensionless quark momentum $x$.
The energy of the nucleon bag is 
\begin{equation}
E_{bag}=3\frac{\Omega_q}{R}-\frac{Z}{R}+\frac{4}{3}\pi R^3 B,
\label{ebag0}
\end{equation}
where $Z$ is a parameter accounting for the zero-point energy and
$B$ is the bag constant. After subtracting spurious
center-of-mass motion inside the bag, the effective mass of
the nucleon bag at rest is given as \cite{jj}
\begin{equation}
M_N^*=\sqrt{E_{bag}^2-\langle p_{cm}^2\rangle},
\end{equation}
with $\langle p_{cm}^2\rangle = \sum_q 
\langle p_q^2\rangle\equiv 3 (x/R)^2$.
The bag radius $R$ is then obtained through
\begin{equation}
\frac{\partial M_N^*}{\partial R}=0.
\label{rr}
\end{equation}

The total energy density of nuclear matter at baryon density
$\rho_B$ is given in the usual form as
\begin{equation}
\epsilon=\frac{\gamma}{(2\pi)^3}\int ^{k_F} d^3 k \sqrt{k^2+{M_N^*}^2}
+\frac{g_\omega^2 \rho_B^2}{2 m_\omega^2}+\frac{1}{2}m_\sigma^2 \sigma^2,
\end{equation}
where $\gamma=4$ is the spin--isospin degeneracy factor
for nuclear matter. The vector mean field $\omega$ is determined through
\begin{equation}
\omega=\frac{g_\omega \rho_B}{m_\omega^2},
\end{equation}
where $g_\omega=3 g_\omega^q$.
Finally, the scalar mean field $\sigma$ is fixed by 
\begin{equation}
\frac{\partial \epsilon}{\partial \sigma}=0.
\label{sig}
\end{equation}
The scalar and vector couplings $g_\sigma^q$ and $g_\omega^q$ 
are fitted to the saturation density and binding energy for nuclear 
matter. For a given baryon density, $x,$ $R,$ and $\sigma$ are 
calculated from the equations (\ref{x}), (\ref{rr}), and (\ref{sig})
respectively. 

In the simple QMC model \cite{qmc,saito}, the bag constant $B$ is taken
as $B_0$ corresponding to the bag parameter for a free nucleon.
The medium effects are taken into account in the modified QMC model
\cite{jj}. To include the finite-temperature effects for nuclear matter,
we choose a specific ansatz for the medium-dependent bag parameter
\cite{jj},
\begin{equation}
B=B_0 \exp(-\frac{4 g_\sigma^B \sigma}{M_N}),
\label{bagc}
\end{equation}
with $g_\sigma^B$ as an additional parameter.
For a fixed set of bag parameters as taken in the ref. \cite{jj},
we now generalize the modified QMC model to study nuclear matter
at finite temperatures.

\section{Nuclear Matter at finite temperatures}

At finite temperatures, the quarks inside the bag can be thermally
excited to higher angular momentum states. However, for simplicity,
we shall still assume the bag describing the nucleon to be spherical
with radius $R$ which is now temperature dependent.

Here the single-particle quark and antiquark energies in units of
$R^{-1}$ are given as
\begin{equation}
\epsilon_{\pm}^{n\kappa}=\Omega^{n\kappa}\pm g_\omega^q\omega R,
\end{equation}
where
\begin{equation}
\Omega^{n\kappa}=(x^2_{n\kappa}+R^2{m^*_q}^2)^{1/2}
\end{equation}
$m^*_q=m^0_q-g_\sigma^q\sigma$ is the effective quark mass.
The boundary condition at the bag surface, given by
\begin{equation}
i\gamma\cdot n \psi_q^{n\kappa}=\psi_q^{n\kappa},
\end{equation}
now determines the quark momentum $x_{n\kappa}$ in the state
characterized by specific values of $n$ and $\kappa$.
The total energy from the quarks and antiquarks is 
\begin{equation}
E_{tot}=3\sum_{n,\kappa}\frac{\Omega^{n\kappa}}{R}\Big[
\frac{1}{e^{(\epsilon_+^{n\kappa}/R-\mu_q)/T}+1}+
\frac{1}{e^{(\epsilon_-^{n\kappa}/R+\mu_q)/T}+1}\Big],
\end{equation}
and the bag energy now becomes
\begin{equation}
E_{bag}=E_{tot}-\frac{Z}{R}+\frac{4\pi}{3}R^3B,
\end{equation}
which reduces to eq.~(\ref{ebag0}) at zero temperature.
The spurious center-of-mass motion in the bag is subtracted to
obtain the effective nucleon mass
\begin{equation}
M_N^*=(E_{bag}^2-\langle p_{cm}^2\rangle)^{1/2}, \quad\quad
\langle p_{cm}^2\rangle = \frac{\langle x^2\rangle}{R^2},
\end{equation}
where
\begin{equation}
\langle x^2\rangle=
3\sum_{n,\kappa}x_{n\kappa}^2\Big[
\frac{1}{e^{(\epsilon_+^{n\kappa}/R-\mu_q)/T}+1}+
\frac{1}{e^{(\epsilon_-^{n\kappa}/R+\mu_q)/T}+1}\Big]
\end{equation}
is the spurious center-of-mass average momentum squared \cite{jj}.
This is here written in terms of the sum of the quark and the
antiquark distributions since the center-of-mass motion does not 
distinguish between a quark and an antiquark.

The quark chemical potential $\mu_q$, assuming that there are
three quarks in the nucleon bag, is determined through
\begin{equation}
n_q=3=3\sum_{n,\kappa}\Big[
\frac{1}{e^{(\epsilon_+^{n\kappa}/R-\mu_q)/T}+1}-
\frac{1}{e^{(\epsilon_-^{n\kappa}/R+\mu_q)/T}+1}\Big].
\label{muq}
\end{equation}
The temperature-dependent bag radius $R$ is obtained 
in the same way as was done at zero temperature
using eq.~(\ref{rr}).

The total energy density at finite temperature $T$ and at finite 
baryon density $\rho_B$ is
\begin{equation}
\epsilon=\frac{\gamma}{(2\pi)^3}\int d^3 k \sqrt{\vec k
+{M_N^*}^2}(f_B+\bar f_B)+\frac{g_\omega^2}{2m_\omega^2}\rho_B^2+
\frac{1}{2}{m_\sigma^2}\sigma^2,
\end{equation}
where $f_B$ and $\bar f_B$ are the thermal distribution functions
for the baryons and antibaryons,
\begin{equation}
f_B=\frac{1}{e^{(\epsilon^*(\vec k)-\mu_B^*)/T}+1} \quad{\rm and}\quad
\bar f_B=\frac{1}{e^{(\epsilon^*(\vec k)+\mu_B^*)/T}+1}
\end{equation}
with
$\epsilon^*(\vec k)=(\vec k^2+{M^*_N}^2)^{1/2}$ the effective
nucleon energy and $\mu_B^*=\mu_B-g_\omega \omega$ the effective
baryon chemical potential.

The thermodynamic grand potential is
\begin{eqnarray}
\Omega&=&\epsilon -TS -\mu_B\rho_B\nonumber\\
&=& \frac{\gamma}{(2\pi)^3}\int d^3 k \epsilon^*(k)
(f_B+\bar f_B)+\frac{g_\omega^2}{2m_\omega^2}\rho_B^2
+\frac{1}{2}m_\sigma^2\sigma^2-TS-\mu_B\rho_B,
\label{therm}
\end{eqnarray}
with the entropy density
\begin{equation}
S= -\frac{\gamma}{(2\pi)^3}\int d^3 k \Big[ f_B\ln f_B+(1-f_B)\ln (1-f_B)+
\bar f_B\ln \bar f_B+(1-\bar f_B)\ln (1-\bar f_B)\Big]
\end{equation}
and the baryon density
\begin{equation}
\rho_B= \frac{\gamma}{(2\pi)^3}\int d^3 k (f_B-\bar f_B).
\end{equation}

The pressure is the negative of $\Omega,$ and, 
after integration by parts,
reduces to the familiar expression \cite{eisen}
\begin{equation}
P= \frac{\gamma}{3}\frac{1}{(2\pi)^3}\int d^3 k 
\frac{{\vec k}^2}{\epsilon^*(k)} (f_B+\bar f_B)
+\frac{g_\omega^2}{2m_\omega^2}\rho_B^2
-\frac{1}{2}m_\sigma^2\sigma^2.
\end{equation}

At finite temperatures and for a given $\mu_B$, 
the effective nucleon mass is known for given values of the meson fields
once the bag radius $R$ and the quark chemical potential $\mu_q$ are 
calculated by using eqs.~(\ref{rr})
and (\ref{muq}) respectively. The mean field $\sigma$ 
is determined through the minimization of the thermodynamic potential 
and using the self-consistency condition 
\begin{equation}
\omega=\frac{g_\omega}{m_\omega^2}
\frac{\gamma}{(2\pi)^3}\int d^3 k (f_B-\bar f_B).
\label{omg}
\end{equation}

\section{Results and Discussions}

We now proceed with the finite--temperature calculations for nuclear matter. 
As already mentioned, we assume the density--dependent bag constant
to be of the form given by eq.~(\ref{bagc}) with $g_\sigma^B$ as a parameter. 
Further, we take the bag parameters for the free nucleon as $B_0^{1/4}$=
188.1 MeV and Z=2.03 \cite{jj} and the current quark mass $m_q^0$ as zero. 
For $g_\sigma^q=1$, the values of the vector meson
coupling and the parameter $g_\sigma^B$, as
fitted from the saturation properties of nuclear matter \cite{jj},
are given as $g_\omega^2/4\pi=5.24$ and $(g_\sigma^B)^2/4\pi=3.69$.

For specific values of the temperature and $\mu_B$, the thermodynamic
potential is given in terms of the effective nucleon mass which depends
on the bag radius $R$, the quark chemical potential $\mu_q,$
and the mean fields $\sigma$ and $\omega$. For given values of
$\sigma$ and $\omega$, the bag radius and the quark chemical potential
are determined using the conditions (\ref{rr}) and (\ref{muq})
respectively. We then determine the value of $\sigma$ by minimizing 
the thermodynamic potential $\Omega$, with $\omega$ determined 
from the self-consistency condition of eq.~(\ref{omg}).   

For given values of the temperature and $\mu_B$, we calculate the
different thermodynamic quantities along with the bag properties, 
which are now temperature dependent. In fig.~1, we plot the pressure
as a function of the baryon density $\rho_B$ for different values of 
the temperature. For a given $\rho_B$, pressure has the usual trend 
of increasing with temperature \cite{furnst}. The pressure for $\rho_B=0$
becomes nonzero at and above a temperature of 200 MeV. This has contributions
arising from the thermal distribution functions for the baryons and
antibaryons as well as from a nonzero value for the sigma field.
The scalar sigma field attaining a nonzero value was also observed
for nuclear matter in the Walecka model \cite{furnst}, which had led to a 
sharp fall in the effective nucleon mass between temperatures
of 150 MeV and 200 MeV. This rapid fall of $M_N^*$ with increasing
temperature resembles a phase transition when the system becomes a 
dilute gas of baryons in a sea of baryon--antibaryon pairs.

The density- and temperature-dependence of the baryon effective mass is
shown in fig.~2. The value of $M_N^*$ here increases monotonically with
temperature. This is in contrast to the calculation of hot nuclear 
matter using Walecka model \cite{furnst}, where, as the temperature
is increased, $M_N^*$ first rises and then falls rapidly
for $T\simeq 200$ MeV. We do not encounter any change in the trend 
here even up to a temperature of 250 MeV, even though similar to 
earlier calculation, the sigma field becomes finite at a temperature
of 200 MeV. The reason for this is that the sigma field here is not
strong as in Ref. \cite{furnst}. Also, there is a significant 
contribution to the effective mass arising from the thermal excitations
of the quarks inside the nucleon bag. This adds to the mass of the nucleon
whereas the nucleon mass decreases with increase in the value of the
sigma field. The contribution of the former, which was absent in the 
Walecka model calculations \cite{furnst}, appears to dominate over
the contribution from the nonzero value of $\sigma$ and the net effect
is a rise of the effective nucleon mass in the present case. Hence, 
although in the earlier case \cite{furnst} the effective mass is 
identical to the free nucleon mass at zero density and nonzero 
temperatures before the phase transition
takes place, $M_N^*$ in our calculations has a higher value as
compared to the nucleon mass in vacuum. A similar increase was also
observed earlier for the $N$ and $\Delta$ masses for nonzero temperatures
using the thermal skyrmions \cite{dey}.

We then look at the dependence of the entropy density on the density
and temperature as shown in fig.~3.  This has a nonzero value even at
vanishing baryon density at and above a temperature of 200 MeV, with 
contributions from the nonzero value of the sigma field.  A similar 
behavior was observed for the entropy density in the Walecka model
calculations \cite{furnst}. The change in the behavior of the entropy
density in the earlier case was rather abrupt indicating a sharp phase
transition. The present calculations show a softening of the phase 
transition.

The mean field $\sigma$ as obtained through the minimization of
the thermodynamic potential is plotted as a function
of the baryon density for various temperatures in fig.~4. 
The sigma field becomes nonzero at a temperature of 200 MeV,
indicating a phase transition to a system of baryon--antibaryon pairs
at very low densities as mentioned earlier.
The value of $\sigma$ decreases with temperature at higher densities.
A similar behavior was also observed earlier at temperatures before
the phase transition took place \cite{furnst}, and had the natural 
consequence that the effective nucleon mass grows with temperature.
However, at low densities and high temperatures, $\sigma$ increases
with temperature thus reducing the effective mass in the earlier case.
Even though the behavior of sigma is similar, the effective mass
here is dominated by the thermal excitations of the quarks which 
leads to an increase in the effective mass with temperature.

We also look at some other properties of the nucleon bag at finite
temperatures. The bag constant, as shown in fig.~5, grows with 
temperature for densities greater than around 0.1 fm$^{-3}$. This is
clear from the form of $B,$ chosen as in (\ref{bagc}), since the mean 
field $\sigma$ decreases with temperature. The bag constant at zero
density is seen to be identical to the free nucleon bag constant for 
temperatures before the phase transition. However, at temperatures
at and above 200 MeV, it starts decreasing with temperature.
The temperature- and density-dependence of the bag radius, in units 
of the free nucleon radius $R_0$=0.6 fermis, is shown in fig.~6. 
The nucleon bag in this model shrinks with increasing temperature. 
Such a difficulty of obtaining a baryon that swells with temperature
was also encountered using the thermal skyrmion \cite{dey,nowak}. A 
possible solution for the difficulty was suggested in the earlier case
\cite{dey,nowak} whereby the thermal skyrmion exists in a bath of pions, 
which effectively increase its radius.

To summarize, we have studied nuclear matter at finite temperatures
using the QMC model and taking into account the density--dependence of the
bag constant. The mean fields $\sigma$ and $\omega$ were determined 
through minimization of the thermodynamic potential and the temperature
dependent bag radius was calculated by minimizing the effective mass
of the nucleon bag. The scalar mean field $\sigma$ at zero density
attains a nonzero value at a temperature of 200 MeV similar to the
Walecka model calculations for the nuclear matter \cite{furnst}, 
which is indicative of a phase transition to a system with 
baryon--antibaryon pairs. However, there is a softening in the
phase transition here as compared to the earlier calculations. 
This is because the thermal contributions from the quarks,
which were absent in Ref. \cite{furnst}, are dominant here and
lead to a rise of the effective nucleon mass with temperature.
The nucleon bag in the present calculation shrinks in size with 
increase in the temperature. The nucleon mass at finite temperatures
and zero baryon density as calculated here is appreciably different 
from the nucleon mass in vacuum.

\section{Acknowledgments}

The authors would like to acknowledge many useful discussions with 
Prof. Henry Jaqaman, Dr. J. Reinhardt, Dr. M. Belkacem, 
Dr. S. Schramm and Dr. H. Mishra. J.M.E.~wishes to thank Professor Walter
Greiner and the Institute for Theoretical Physics at the University of 
Frankfurt for their kind hospitality, and to acknowledge support from
the Deutsche Forschungsgemeinschaft and the Ne'eman Chair in Theoretical
Nuclear Physics at Tel Aviv University. A.M. and P.K.P. would like
to thank the Alexander von Humboldt-Stiftung for financial support and
the Institut f\"ur Theoretische Physik for hospitality.

\vfil
\eject
\centerline{\bf Figures Captions}

\noindent Fig.~1. The pressure for nuclear matter as
a function of the baryon density $\rho_B$ for various values of 
temperature showing an increase with temperature.
\hfil\break

\noindent Fig.~2. The effective nucleon mass for nuclear matter as a 
function of the baryon density $\rho_B$ for different values
of the temperature, which is seen to rise monotonically with 
temperature. \hfil\break

\noindent Fig.~3. The entropy density versus the density for
various temperatures. \hfil\break

\noindent Fig.~4. The mean field $\sigma$ as a function of density
for different temperatures.
\hfil\break

\noindent Fig.~5. The bag constant versus the density for different
temperatures. It grows with temperature, except for very low
densities and temperatures at and above 200 MeV. \hfil\break

\noindent Fig.~6. The bag radius as a function of density at
various temperatures; the nucleon bag size shrinks with temperature.
\vfil
\eject

\begin{figure}
\epsfbox{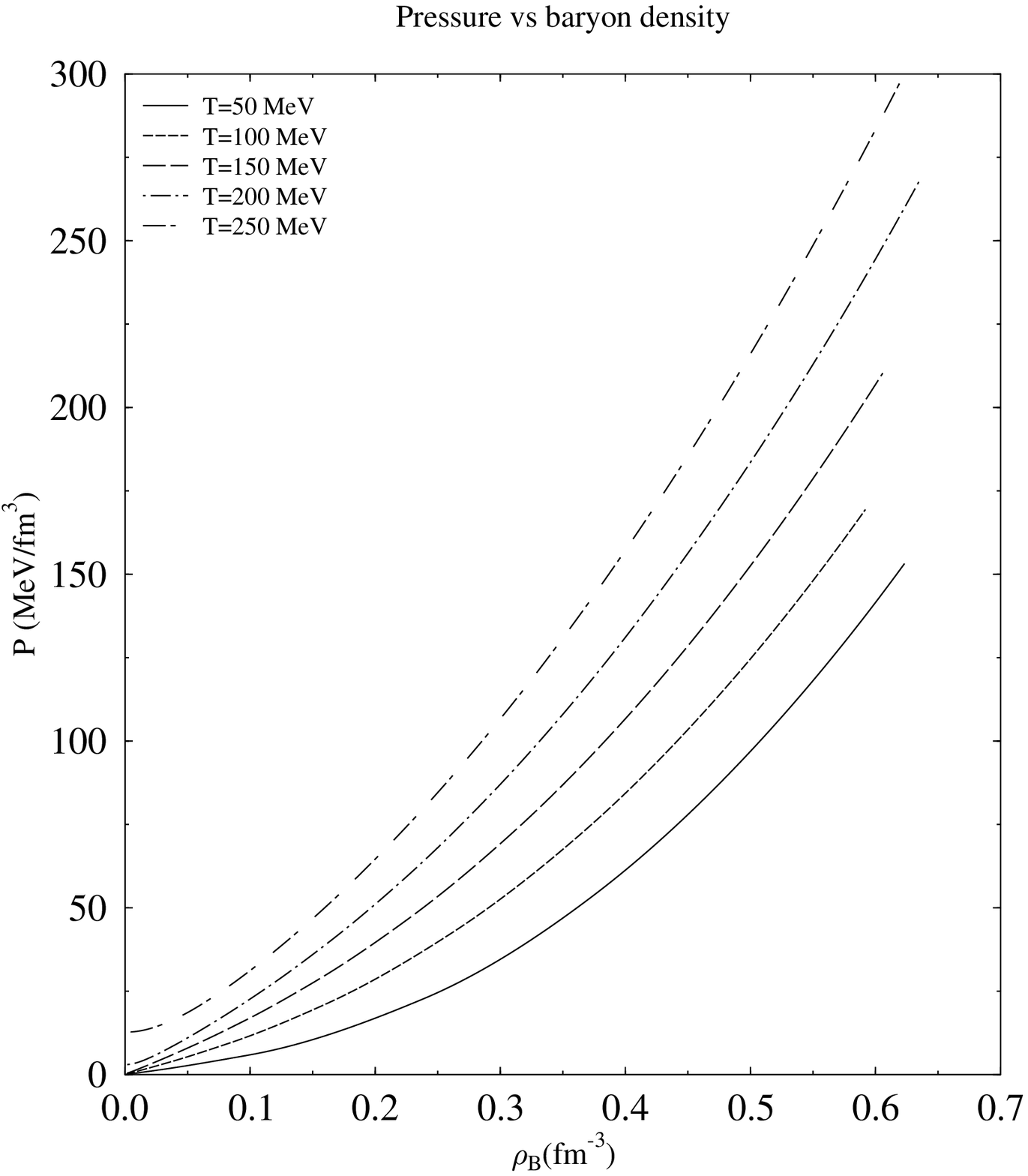}
\end{figure}
\begin{figure}
\epsfbox{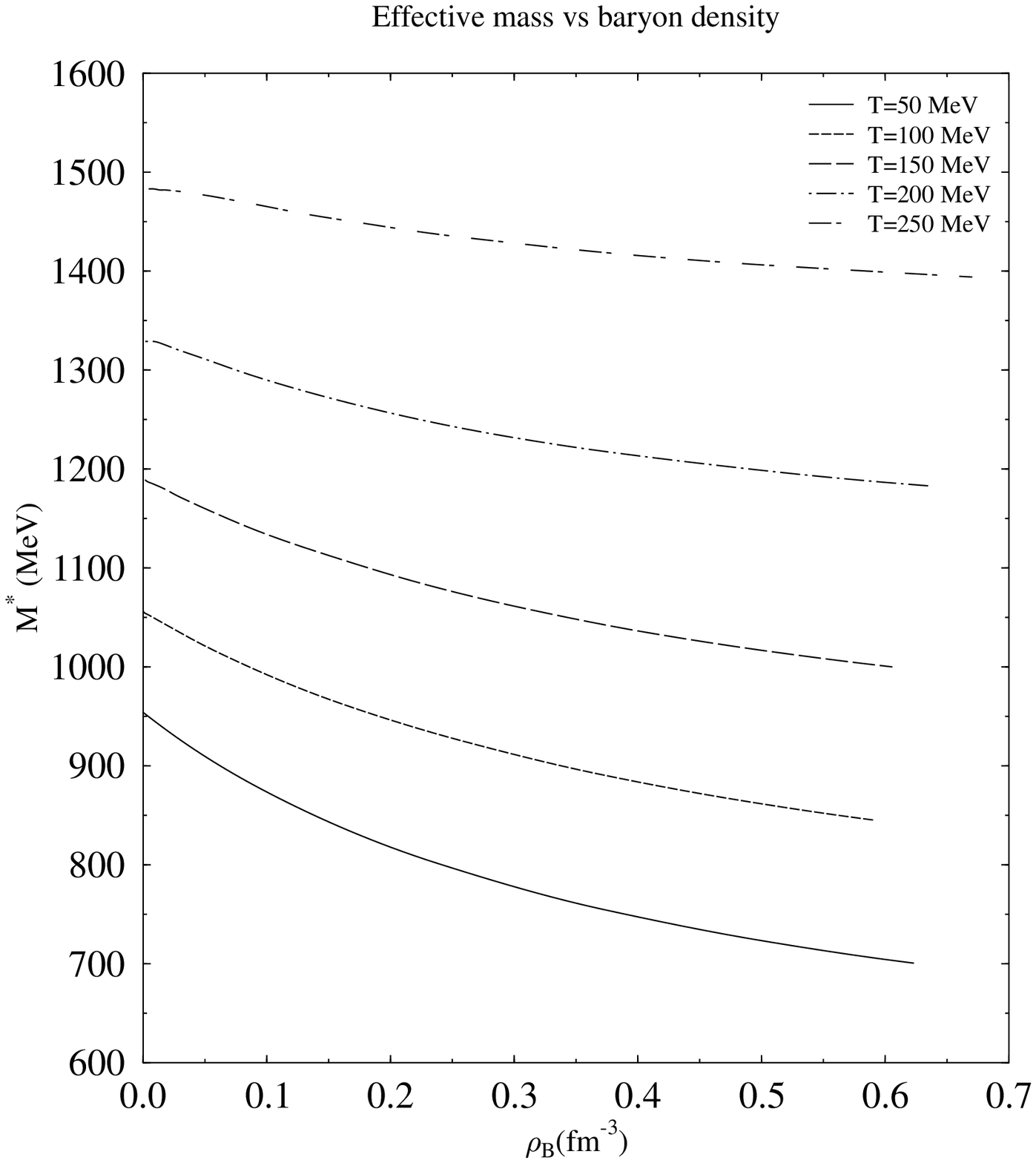}
\end{figure}
\begin{figure}
\epsfbox{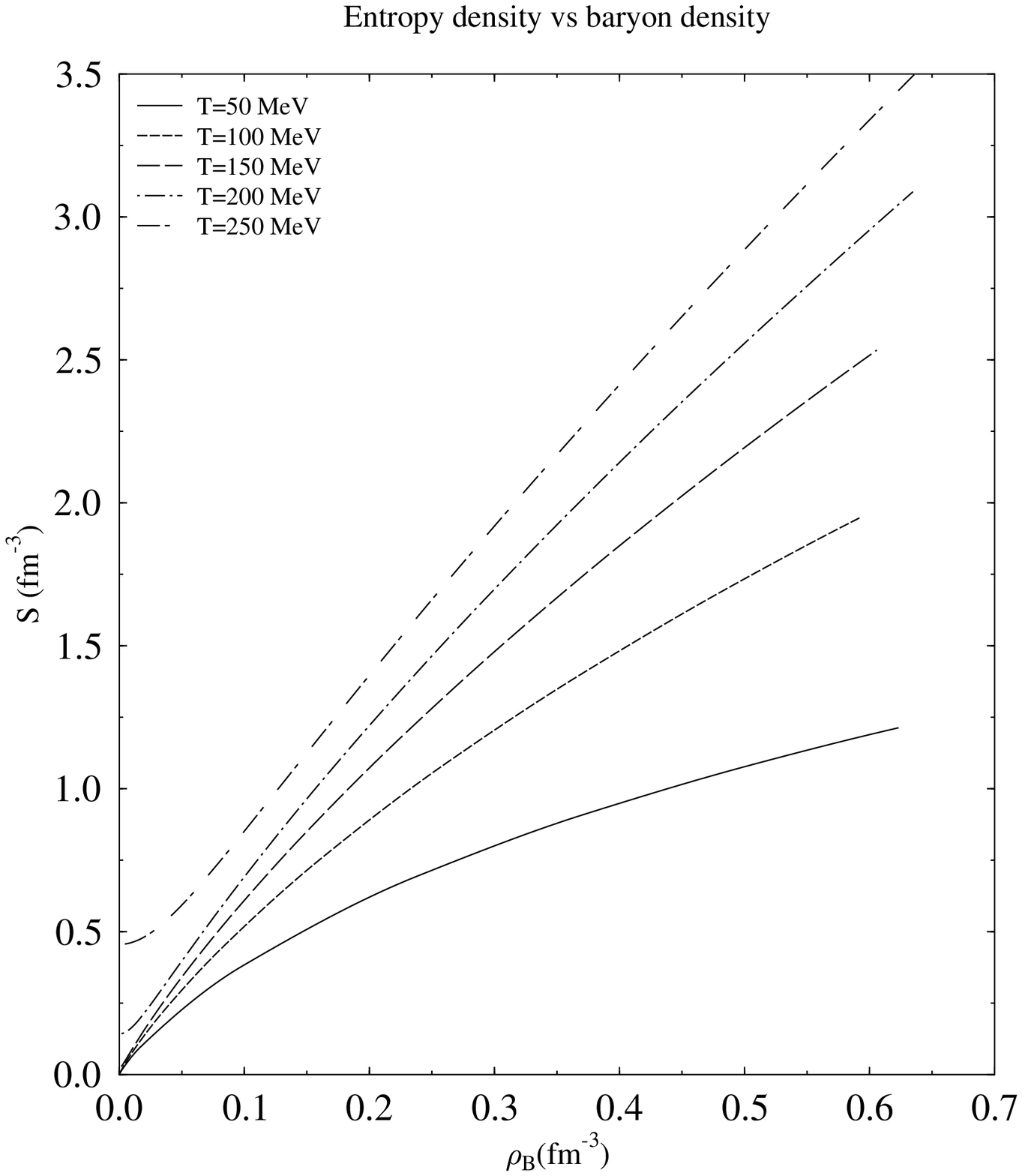}
\end{figure}
\begin{figure}
\epsfbox{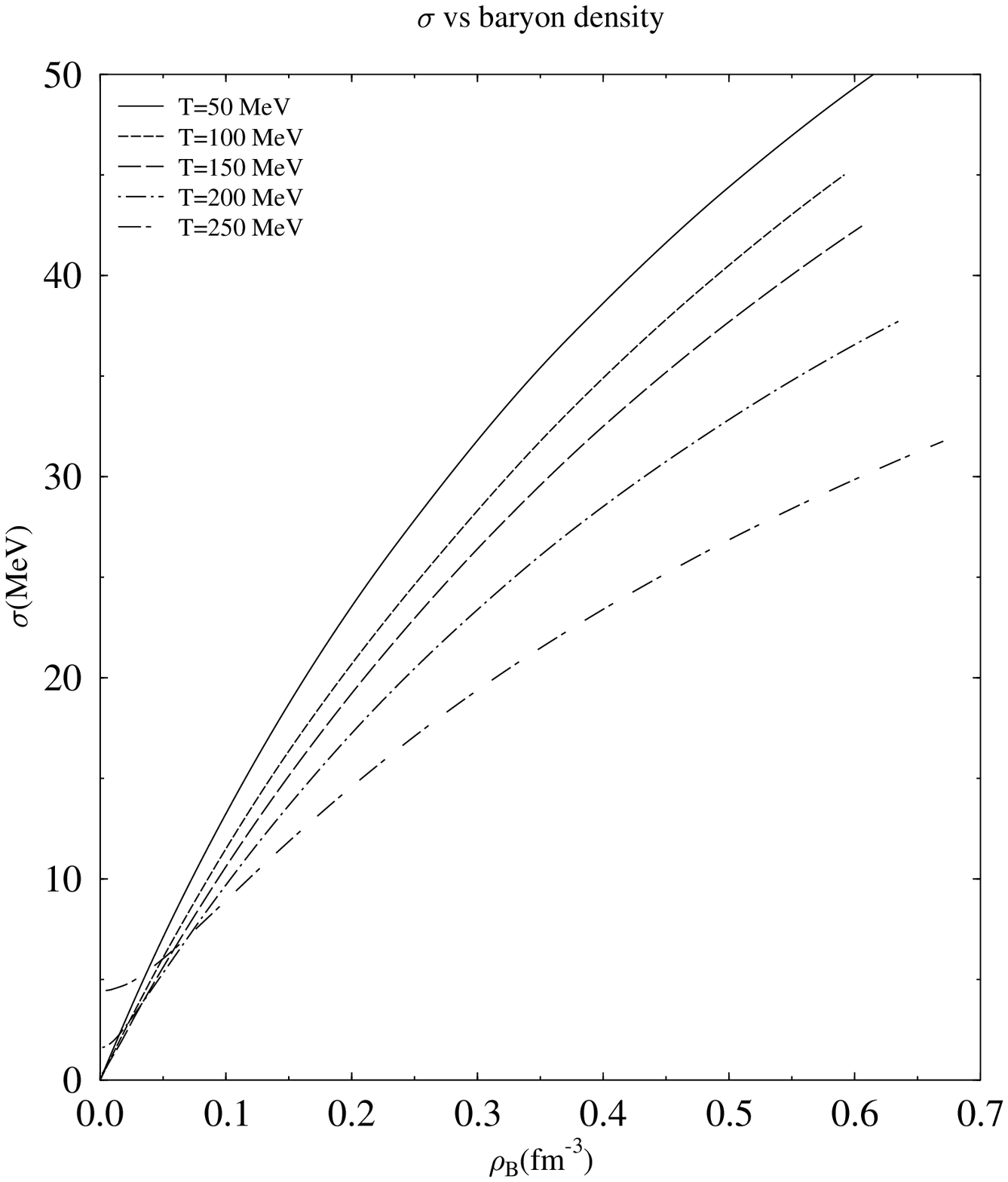}
\end{figure}
\begin{figure}
\epsfbox{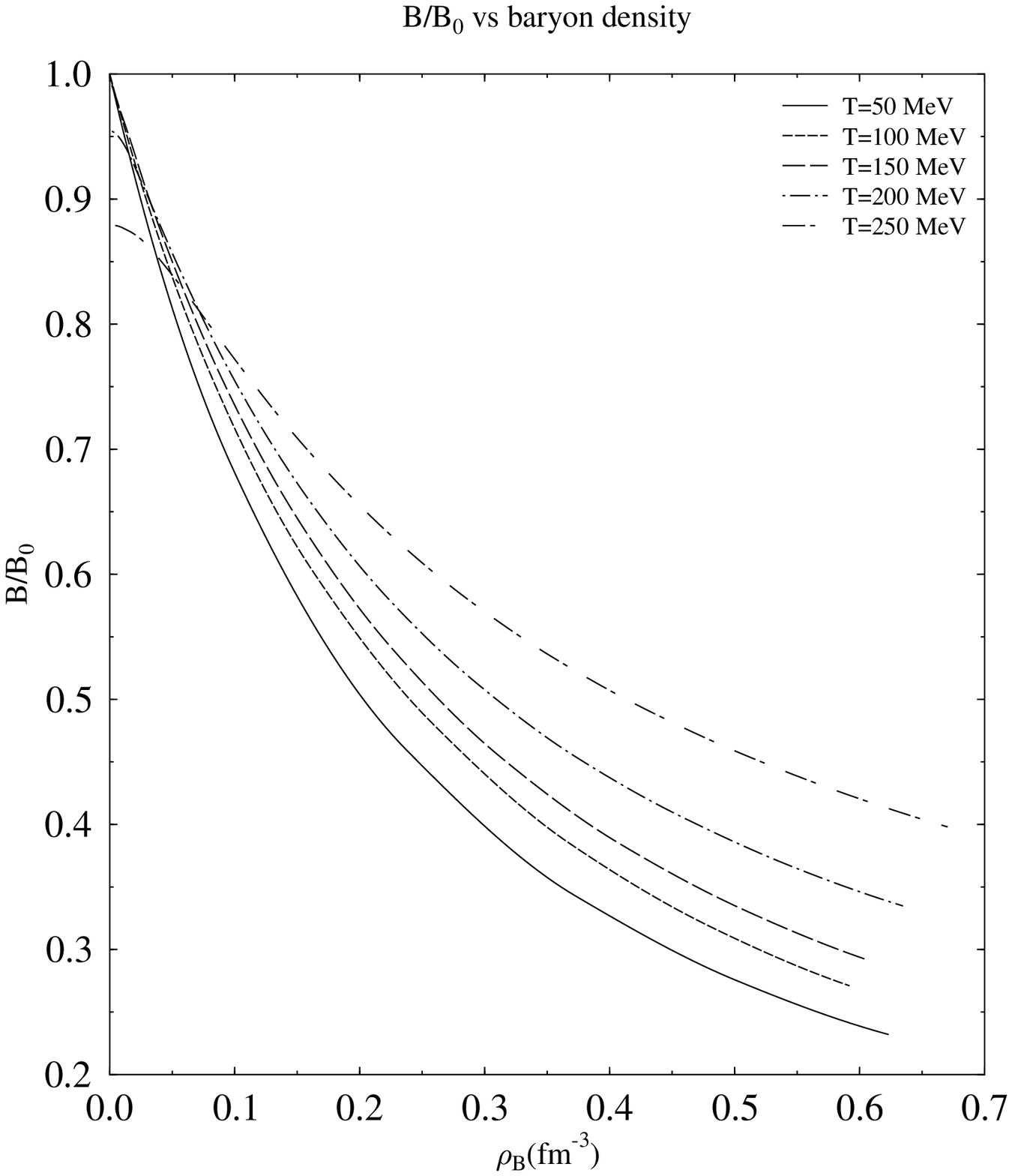}
\end{figure}
\begin{figure}
\epsfbox{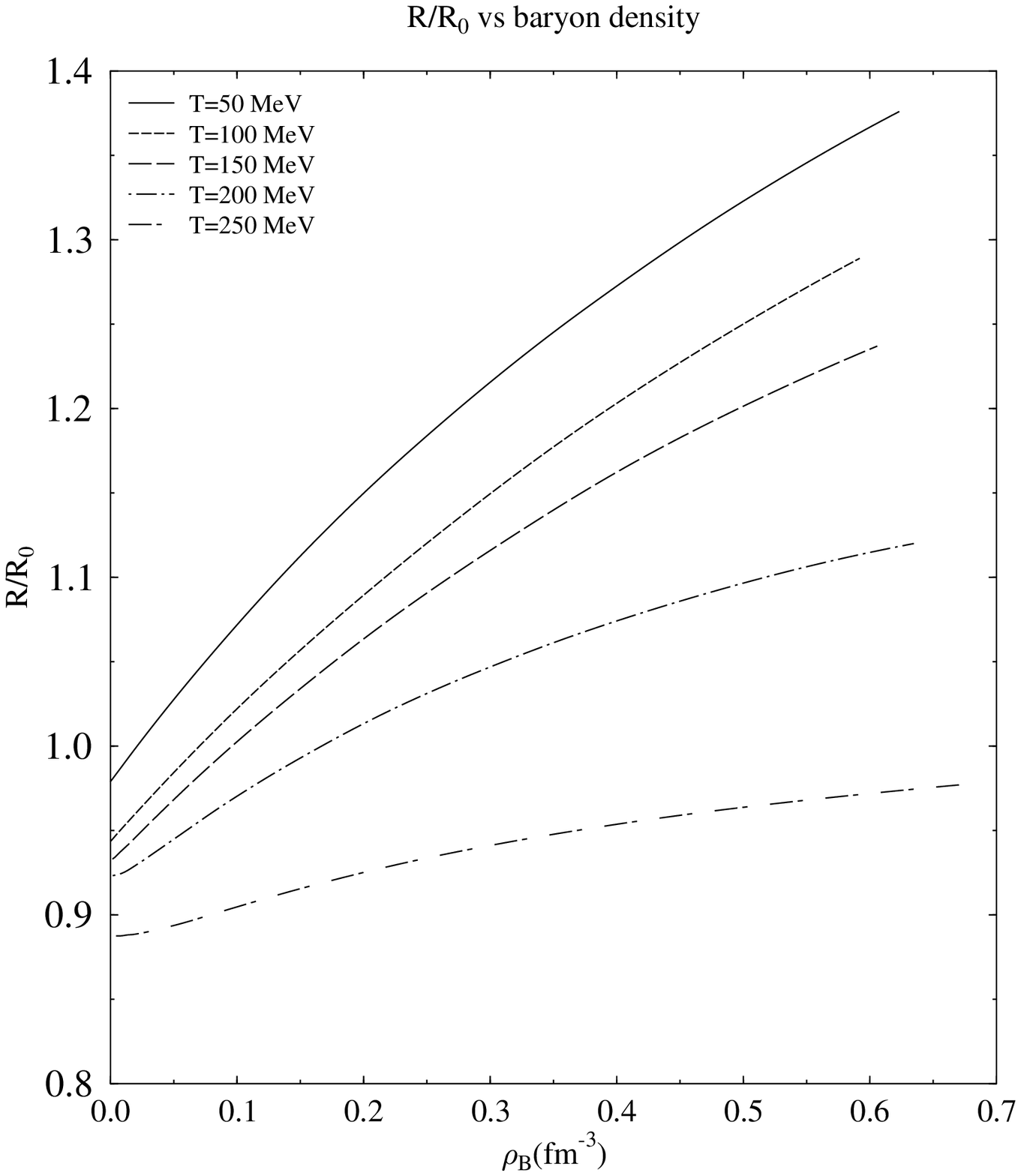}
\end{figure}
\end{document}